\documentclass[12pt, a4paper, twocolumn]{article}

\usepackage{amsmath,amssymb,amsfonts}
\usepackage{amsthm}
\usepackage{graphicx} 
\usepackage{acronym}
\usepackage{manyfoot}
\usepackage{xcolor}
\usepackage{physics}
\usepackage{layouts}
\usepackage{hyperref}
\usepackage[capitalise]{cleveref}
\usepackage{subcaption}
\usepackage{authblk}

\newcommand{\ve}[1]{\ensuremath{\mathbf{#1}}} 
\DeclareMathOperator{\de}{\mathrm{d}}

\acrodef{ml}[ML]{Machine Learning}
\acrodef{dm}[DM]{Diffusion Model}
\acrodef{gan}[GAN]{Generative Adversarial Network}
\acrodef{qgan}[QGAN]{Quantum Generative Adversarial Network}
\acrodef{qml}[QML]{Quantum Machine Learning}
\acrodef{vae}[VAE]{Variational Auto Encoder}
\acrodef{ddpm}[DDPM]{Denoising Diffusion Probabilistic Model}
\acrodef{qddpm}[Q-DDPM]{Quantum Denoising Diffusion Probabilistic Model}
\acrodef{nn}[NN]{neural network}
\acrodef{pqc}[PQC]{Parametrized Quantum Circuit}
\acrodef{arm}[ARM]{Autoregressive Model}
\acrodef{tom}[TOM]{transition operation matrix}
\acrodef{tem}[TEM]{transition effect matrix}
\acrodef{povm}[POVM]{positive operator valued measure}
\acrodef{sde}[Score SDE]{Score Stochastic Differential Equation}
\acrodef{sgm}[SGM]{Score-based Generative Model}
\acrodef{sse}[SSE]{Stochastic Schr\"odinger Equation}
\acrodef{qndg}[QNDG]{quantum noise driven generative}
\acrodef{qpu}[QPU]{quantum processing unit}
\acrodef{qnn}[QNN]{Quantum Neural Network}
\acrodef{cqgdm}[CQGDM]{Classical-Quantum Generative Diffusion Model}
\acrodef{qcgdm}[QCGDM]{Quantum-Classical Generative Diffusion Model}
\acrodef{qqgdm}[QQGDM]{Quantum-Quantum Generative Diffusion Model}
\acrodef{qndgdm}[QNDGDM]{quantum-noise-driven generative diffusion model}
\acrodef{nisq}[NISQ]{Noisy Intermediate-Scale Quantum}
\acrodef{tom}[TOM]{transition operation matrix}
\acrodef{tem}[TEM]{transition effect matrix}
\acrodef{nlp}[NLP]{Natural Language Processing}
\acrodef{kl}[KL]{Kullback-Leibler}
\acrodefplural{tom}[TOMs]{transition operation matrices}
\acrodefplural{tem}[TEMs]{transition effect matrices}
\begin{document}

\title{Quantum-Noise-Driven Generative Diffusion Models}

\author[1]{Marco Parigi}
\author[1,2]{Stefano Martina}
\author[1,2]{Filippo Caruso}

\affil[1]{Department of Physics and Astronomy, University of Florence, Via Sansone 1, Sesto Fiorentino, 50019, Florence, Italy.}
\affil[2]{LENS - European Laboratory for Non-Linear Spectroscopy, University of Florence, Via Nello Carrara 1, Sesto Fiorentino, 50019, Florence, Italy.}

\date{marco.parigi@unifi.it, stefano.martina@unifi.it, filippo.caruso@unifi.it}

\maketitle

\begin{abstract}
Generative models realized with machine learning techniques are powerful tools to infer complex and unknown data distributions from a finite number of training samples in order to produce new synthetic data. Diffusion models are an emerging framework that have recently overcome generative adversarial networks in creating high-quality images.
Here, is proposed and discussed the quantum generalization of diffusion models, i.e., three quantum-noise-driven generative diffusion models that could be experimentally tested on real quantum systems. The idea is to harness unique quantum features, in particular the non-trivial interplay among coherence, entanglement and noise that the currently available noisy quantum processors do unavoidably suffer from, in order to overcome the main computational burdens of classical diffusion models during inference. Hence, the suggestion is to exploit quantum noise not as an issue to be detected and solved but instead as a beneficial key ingredient to generate complex probability distributions from which a quantum processor might sample more efficiently than a classical one. We also include three examples of the numerical simulations for the proposed approaches. The results are expected to pave the way for new quantum-inspired or quantum-based generative diffusion algorithms addressing tasks as data generation with widespread real-world applications.
\end{abstract}

\section{Introduction}

\begin{figure*}[!ht]
\begin{center}
\includegraphics[width=\linewidth]{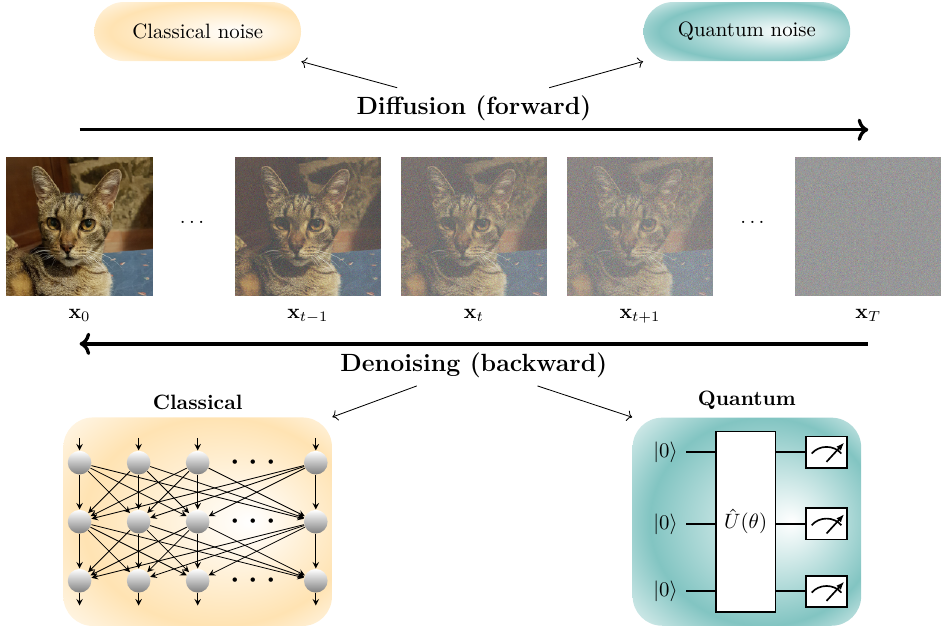}
\caption{Depiction of the diffusion (from left to right) and denoising (from right to left) processes within a diffusion probabilistic model framework. The original image $\ve{x}_0$ sampled from the unknown data distribution $p(\ve{x}_0)$ is progressively perturbed ($t \rightarrow t+1$) by adding noise to obtain a latent variable $\ve{x}_T$ from a known and tractable distribution where the information is completely destroyed. In our framework the diffusion process can be implemented with a classical or a quantum stochastic dynamics. The denoising process is trained to approximate the structure of the data distribution in order to generate new samples. The latter is implemented step by step, using a classical (on the left in orange) or quantum (on the right in green) parameterized model $\hat{U}(\theta)$ in order to approximate the backward mapping. The standard diffusion models implement both the diffusion and the denoising processes in a classical framework. We propose three different new approaches for the other cases: i) classical diffusion and quantum denoising (\acs{cqgdm}); ii) quantum diffusion and classical denoising (\acs{qcgdm}); iii) quantum diffusion and quantum denoising (\acs{qqgdm}). A similar picture can be applied to time series.}
\label{fig:fig1}
\end{center}
\end{figure*}

In \ac{ml}, diffusion probabilistic models, or briefly \acp{dm}, are an emerging class of generative models used to learn an unknown data distribution in order to produce new data samples. They has been proposed for the first time by Sohl-Dickstein et al.~\cite{Sohl2015} and take inspiration from diffusion phenomena of non-equilibrium statistical physics. The underlying core idea of the \acp{dm} is to gradually and slowly destroy the information encoded into the data distribution until it became fully noisy, and then learn how to restore the corrupted information in order to generate new synthetic data. More precisely, the generic structure of diffusion models consists of two stages: (i) a \emph{diffusion} (or \emph{forward}) process and (ii) a \emph{denoising} (or \emph{reverse}) process. In the former phase, a training data is progressively perturbed by adding noise, typically Gaussian, until all data information is destroyed. The increasing perturbation of information due to the systematically and progressive injections of noise can be physically understood as if the noise propagates inside the data structure, as shown in \cref{fig:fig1} from left to right. Let us highlight the fact that in this first stage the training of any \ac{ml} model is not required. In the second phase, the previous diffusive dynamics is slowly reversed in order to restore the initial data information. The goal of this phase is to learn how to remove noise correctly and produce new data starting from uninformative noise samples as in \cref{fig:fig1} from right to left. In contrast to the forward diffusion process, the noise extraction---and as a result the data information retrieval---is implemented training a \ac{ml} model typically based on a so-called U-Net \ac{nn} architecture~\cite{Ronneberger2015}. In detail, U-Net models are structured in a succession of convolutional layers followed by an equal number of deconvolutional layers where each deconvolution takes as input the output of the previous deconvolution and also the copy of the output of the corresponding convolutional layer in reverse order. The procedure described above allows \acp{dm} to succesfully address the main complication in the design of probabilistic models, i.e., being \emph{tractable} and \emph{flexible} at the same time~\cite{Sohl2015,Ho2020}. In fact, alternatively to \acp{dm} there are other generative probabilistic models, for instance, \acp{arm} that are generally tractable but not flexible, or \acp{vae}~\cite{Kingma2019} and \acp{gan}~\cite{Goodfellow2020} that are flexible but not tractable.

Diffusion models find use in computer vision for several image processing tasks~\cite{Croitoru2023}, such as, inpainting~\cite{Lugmayr_2022_CVPR}, super-resolution~\cite{Saharia2023}, image-to-image translation~\cite{Saharia2022palette}, and image generation~\cite{Rombach_2022_CVPR,ramesh2022hierarchical,saharia2022photorealistic}. They are also successfully adopted in several applications, for instance: Stable diffusion~\cite{stableDiffusion} that is an open source model for high resolution image syntesis~\cite{Rombach_2022_CVPR}; DALL-E 2 that is a platform implemented by OpenAI~\cite{dalle2} to generate photorealistic images from text prompts~\cite{ramesh2022hierarchical}; Google Imagen~\cite{imagen} that combines \emph{transformer} language models with diffusion models also in the context of text-to-image generation~\cite{saharia2022photorealistic}. Moreover, it has recently been shown that diffusion models perform better than \acp{gan} on image synthesis~\cite{Dhariwal2021}.

Furthermore, diffusion models can also be applied to other contexts, for instance in text generation~\cite{savinov2022stepunrolled,yu2022latent} and time-series related tasks~\cite{lin2023diffusion,Tashiro2021,alcaraz2023diffusionbased}. For instance, time series forecasting is the task of predicting future values from past history and diffusion models can be employed to generate new samples from the forecasting distribution~\cite{Rasul2021,li2022}. Diffusion models can also be used in time series generation, which is a more complex task involving the complete generation of new time-series samples from a certain
distribution~\cite{lim2023regular,adib2023synthetic}.

On the other side, very recently, we are witnessing an increasing interest in quantum technologies. Near-term quantum processors are called \ac{nisq} devices~\cite{preskill2018quantum} and they represent the-state-of-the-art in this context. 
\ac{nisq} computers are engineered with quantum physical systems using different strategies. For instance, a commonly used technology employs superconductive-circuits-based platforms~\cite{devoret2004superconducting,clarke2008superconducting} realized with transmon qubits~\cite{Koch2007charge,Schreier2008suppressing}. This technology is exploited, for instance, by IBM~\cite{ibmQuantum}, Rigetti Computing~\cite{rigetti}, Google~\cite{googleQuantumAI}. Moreover D-Wave~\cite{dwave} exploits superconducting integrated circuits mainly as quantum annealers~\cite{johnson2011quantum}. Xanadu~\cite{xanadu} is instead a company employing photons as information units within the linear optical quantum computing paradigm~\cite{Kok2007linear} to realize their devices. Finally, the quantum computation can be realized directly manipulating the properties of single atoms. For instance, IonQ~\cite{ionq} realizes quantum devices with trapped ions~\cite{Allen2017reconfigurable,Monroe2013scaling}, while Pasqal~\cite{pasqal} and QuEra~\cite{quera} realize analog quantum computers with Rubidium Rydberg neutral atoms held in optical tweezers~\cite{Henriet2020quantumcomputing}.
All the mentioned devices can be in principle integrated in computational pipelines that can involve also classical computation. In this context they can be referred with the term \ac{qpu} that can make some computational task much faster than its classical counterpart (CPU) harnessing the quantum properties of particles at the atomic scale.
The main reason for building a quantum processor is the possibility of exploiting inherent and peculiar resources of quantum mechanical systems such as \emph{superposition}, \emph{coherence} and \emph{entanglement} that, in some cases, allow to perform computational tasks that are impossible or much more difficult via a classic supercomputer~\cite{Nielsen2002,feynman1982simulating}. In particular, the peculiar properties of quantum systems could lead to \emph{quantum speedup} on some tasks compared to their classical counterpart~\cite{deutsch1992,shor1997,grover1996}.

One of the most promising applications of \ac{nisq} devices is represented by \ac{qml} that is a recent interdisciplinary field merging \ac{ml} and quantum computing fields in a way such that data to be processed and/or learning algorithms are quantum~\cite{biamonte2017quantum,wittek2014quantum,schuld2021machine,Nielsen2002}. Indeed, it involves the integration of \ac{ml} techniques and quantum computers in order to process and subsequently analyze/learn the underlying data structure.
\ac{qml} can involve the adoption of classical \ac{ml} methods with quantum data or environments, for instance to analyze noise in quantum devices~\cite{canonici2023machinelearning,Martina_2023_spectroscopy,Martina_2023_non_markovian,martina2022learning} or to control quantum systems~\cite{dalla2022quantum}. Alternatively, \ac{qml} can consider the implementation of novel \ac{ml} techniques using quantum devices, for instance to implement visual tasks~\cite{das2023quantum} or generative models like \ac{qgan}~\cite{Lloyd2018,Dallaire-Demers2018,zoufal2019quantum,braccia2021,braccia2022} that are the quantum implementation of classical \ac{gan} or in \ac{nlp} context to generate text~\cite{karamlou2022quantum}.
In fact, quantum devices are capable of processing information in ways that are different from the classical computation. Thus, the implementation of \ac{qml} models can offer an advantage over the corresponding classical \ac{ml} models~\cite{riste2017demonstration,huang2021information}, especially in the context of generative models~\cite{hibat2024framework}.
However, \ac{nisq} devices are indeed still very noisy and thus they do not perform the ideal (pure) dynamics. Therefore, the system evolution is affected and driven by \emph{quantum noise} due to the undesired interactions with external environment and has to be described by the more general open quantum system formalism~\cite{Caruso2014}.

In order to generalize DMs with quantum computing ideas, a crucial role is played by noise. In classical information theory, noise is usually modeled by the framework of probability theory, and in general via Markovian processes. Accordingly, the main features of classical noise are a linear relationship among successive steps of the dynamics whose evolution depends only on the current state. Formally, noise is represented by a transition matrix that has the properties of \emph{positivity} (non-negative entries) and \emph{completeness} (columns summing to one). In particular, Gaussian noise is a type of random noise that is very often added to the input data of a DM in order to help its learning to generate new data that is as similar as possible to the training data, also in the case when the input is not perfect.

In the quantum domain, noise can be generated also by quantum fluctuations that are typical of quantum systems, hence going much beyond the classical noise sources. Mathematically, quantum noise is described by the more general formalism of \emph{quantum operations} or \emph{quantum maps}~\cite{Caruso2014}, where, for instance, the decoherence is the typical noise affecting the phase coherence among the quantum states. The noise is, in fact, the main enemy to fight in order to build up quantum processors that are  more powerful than the classical counterpart. Indeed, recent works show that the noise can even destroy quantum computation making the dynamic more classically simulable~\cite{aharonov2023polynomial}.
But what about if such noise is not only detrimental for the quantum computation but it is instead actually beneficial for some ML tasks? Even considering the aforementioned detrimental effects, there are theoretical and experimental evidences~\cite{caruso2010,caruso2016fast,dalla2022quantum} that quantum noise can improve the efficiency of information transport and a noisy quantum dynamics can diffuse faster than the noiseless equivalent. Quantum noise might allow, for instance, to generate more complex (due to the presence of entanglement) probability distributions that would be difficult, or even impossible, to express classically and from which is possible to sample more efficiently via a quantum processor than via a classical supercomputer.

Compared to other generative models, classical \acp{dm} require a large number of steps, both for the diffusion and the denoising phases. This means that, when used in data generation, the sampling is computationally expensive because it requires to iterate through all such steps. Inspired by the aforementioned physical systems where the quantum noise accelerates the diffusion,
in this article we therefore introduce and formalize the quantum versions of \acp{dm}, in particularly based on \acp{ddpm} and \acp{sde} in the context of \ac{qml}. More precisely, we propose three potential \acp{qndgdm} that can be both computationally simulated in the NISQ devices and implemented \emph{experimentally} due to the naturally occurring noise effects in open quantum systems. The three algorithms are: i) \acf{cqgdm} in which the forward diffusion process can be implemented in the classical way, while the backward denoising with a \ac{qnn} (that can be either a \ac{pqc} or an hybrid quantum-classical \ac{nn}); (ii) \acf{qcgdm} in which the noise diffusion process can be implemented in a quantum way, while in the denoising process classical NNs are used; (iii) \acf{qqgdm} where both the diffusion and the denoising dynamics can be implemented in a quantum domain.

\section{Results}
\subsection{\acf{cqgdm}}\label{sec:cqgdm}

In this section we propose a model where the diffusion process is classical while the denoising phase is implemented with a quantum dynamics. Moreover, as a result of this setting, the training dataset is necessarily classical, for instance, images, videos, time series, etc.

Formally, given an initial training data $\ve{x}_0$ sampled from a generic and unknown probability distribution $p(\ve{x}_0)$, the procedure consists in a progressive 
destruction of the information encoded in the initial data via a diffusive stochastic process. At the end, the data is degraded to a fully noisy state $\ve{x}_T$ sampled from a classical closed form and tractable \emph{prior} distribution $p(\ve{x}_T)$ that represents the latent space of the model. 
Here, tractable stands for the fact that the distribution can be   computationally calculated. The implementation of this process can be obtained with different ways. 
For instance, in \acp{ddpm} the dynamics of forward diffusive process is implemented by a classical Markov chain~\cite{Sohl2015,Ho2020}, while in \acp{sde} the stochastic evolution is determined by a differential equation~\cite{song2021scorebased}. 
In detail, the former approach considers a discrete-time stochastic process whose evolution, at every step, depends only on the previous state and the transition relies on hand-designed kernels $p(\ve{x}_t|\ve{x}_{t-1})$, $t=1, 2, \dots, T$ (see \cref{sec:classicalMethods} for more details). Alternatively, in \acp{sde} the evolution is a continuous-time process within a close time interval $ t \in [0, T]$ and determined by the stochastic differential equation: $\label{eq:sde} \de \ve{x} = \ve{f}(\ve{x}, t) \de t + g(t) \de \ve{w}$, where $\ve{f}(\ve{x}, t)$ is the drift coefficient, $g(t)$ is the diffusion term, and $\ve{w}$ is the Wiener process (also known as standard Brownian motion) that models the stochastic process~\cite{yang2023diffusion}. The solutions of this equation lead to the tractable prior distribution $p(\ve{x}(T))$.

\begin{figure*}[!ht]
\begin{subfigure}[b]{0.6\textwidth}
\begin{center}
    \includegraphics[width=\linewidth]{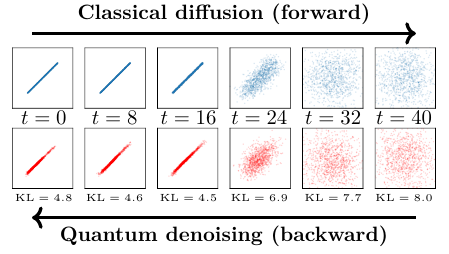}
    \caption{}
    \label{fig:fig2a}
\end{center}
\end{subfigure}%
\begin{subfigure}[b]{0.4\textwidth}
\begin{center}
    \includegraphics[width=0.8\linewidth]{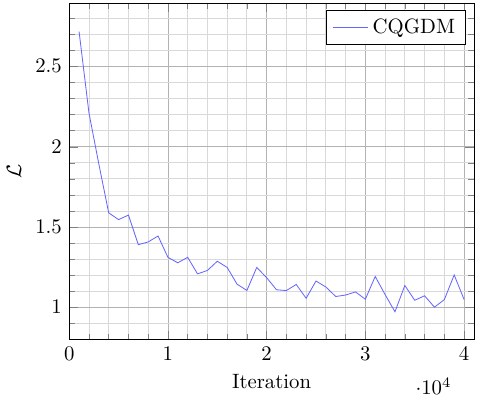}
    \\[0.5cm]
    \caption{}
    \label{fig:fig2b}
\end{center}
\end{subfigure}
\caption{(a) Evolution of the data distribution for a trained simulated \ac{cqgdm} and (b) its \acl{kl} divergence loss function $\mathcal{L}$ during the training, averaged every $1\,000$ iterations. The initial data distribution consists of two-dimensional points distributed in a line segment between $-1$ and $1$. The diffusion process is implemented via a classical diffusion process that transforms the initial data distribution $p(\ve{x}_0)$ at time $t=0$ to the prior $p(\ve{x}_{T})$ that is a normalized Gaussian distribution at the final time $t=40$. Meanwhile, the denoising is implemented via a (noiseless) simulated \ac{pqc} to reconstruct the initial data distribution ($t=0$) from the Gaussian prior ($t=40$). In the top row of (a), we show the forward process (from left to right) for a sample of $1\,000$ points at different discrete time steps $t=0,8,\dots,40$. In the bottom row, we display the denoising (from right to left) of a different sample of $1\,000$ points. Under the figure is reported also the \acl{kl} divergence between the data distribution $p(\ve{x}_0)$ and the reconstructed distribution at the corresponding time.}
\label{fig:fig2}
\end{figure*}

Afterwards, in order to generate new data samples, the objective is to learn how to reverse the diffusion process starting from the prior latent distribution. In case of \acp{ddpm}, calculating $p(\ve{x}_{t-1} | \ve{x}_{t})$ is not computationally tractable and it is classically approximated by a model parameterized with $\theta$ (e.g., a \ac{nn}): $p_\theta(\ve{x}_{t-1} | \ve{x}_{t})$ (see \cref{sec:classicalMethods}). In the case of \ac{sde} models, the quantity to be estimated is $\nabla_\ve{x}\log p_t(\ve{x})$, where $p_t(\ve{x})$ is the density probability of $\ve{x}(t)$~\cite{song2021scorebased}. Here, for either \acp{ddpm} and \ac{sde} diffusion processes, we propose to implement the denoising process with a classical method aided by a \ac{qnn} that can be fully quantum, via a \ac{pqc}, or even a classical-quantum hybrid NN model.
The results of a simulation of this type of algorithm on a dataset composed of 2-dimensional points distributed along a line segment in the interval $[-1,1]$ is shown in \cref{fig:fig2a}. At the best of our knowledge, this is the first implementation of a hybrid classical-quantum diffusion model, and indeed represents a starting point for more in-depth future studies. The model is capable to reconstruct the initial data distribution $p(\ve{x}_0)$ with a good approximation that is quantified by the \ac{kl} divergence $\text{KL}(p(\ve{x}_0)||p_\theta(\ve{x}_{t:T}))$ between the data distribution and the one reconstructed starting from the Gaussian distribution $p(\ve{x}_T)$ until time $t$. In detail, we estimate the \ac{kl} from the $1\,000$ samples using the method described in~\cite{perez2008kullback}. In \cref{fig:fig2b} we show the evolution of the loss during the training of the model averaged every $1\,000$ iterations
 (more details on the model and the implementation in \cref{sec:simulationCQ}).

In this context, the main advantage of using the quantum denoising process instead of the classical one can be the possibility of using the trained quantum model to efficiently generate highly dimensional data (e.g., images) taking advantage of the peculiar quantum mechanical properties, such as quantum superposition and entanglement, to speed up data processing~\cite{lloyd2014quantum,wiebe2012,yao2017}. Indeed, \ac{qpu} devices could be very effective to overcome the main computational burdens of classical diffusion model during this inference process. As shown in \cref{fig:fig4}, the denoising process for \ac{cqgdm} crosses the border between classical and quantum distribution spaces, this could take advantage of the quantum speedup in order to accelerate the training of the model. Moreover, in literature there are evidences of a quantum advantage of using \acp{pqc} instead of deep \acp{nn}. For instance, it is proved that \acp{pqc} outperform classical \acp{nn} in generative tasks~\cite{du2020expressive} and it is shown that \acp{pqc} have an exponential advantage in model size respect to \acp{nn} in function approximation of high dimensional smooth functions~\cite{yu2023provable}. Therefore, it is plausible that we also can take advantage in using \acp{pqc} instead of \acp{nn} in our context.

\subsection{\acf{qcgdm}}\label{sec:qcgdm}

In real experiments quantum systems are never perfectly isolated, but they are easily subjected to noise, e.g., interactions with the environment and imperfect implementations. Accordingly, we propose to physically implement the diffusion process via a noisy quantum dynamics.

In this setting a quantum dataset is considered, i.e., a collection of \emph{quantum} data. Classical information can be embedded into the initial state of a quantum system, allowing to treat classical data as quantum~\cite{schuld2021machine,lloyd2020quantum,Gianani2022experimental}. Even better, we could avoid the encoding of the classical data if we consider quantum data as any result arising from a quantum experiment~\cite{huang2021power} or produced directly by a quantum sensing technology~\cite{huang2021information}. 
Formally, a quantum data is identified with the density operator $\rho$ living in $ \mathfrak{S}(\mathcal{H})$ being the set of non-negative operators with unit trace acting on the elements of the Hilbert space $\mathcal{H}$ where the quantum states live.

We here propose two approaches to implement the diffusion process: (i) \emph{quantum} Markov chains generalizing their classical counterparts~\cite{gudder2008}, and (ii) \ac{sse}~\cite{davies1970operational,wiseman1996quantum,breuer2002theory,rivas2012open} modelling the dynamics of an \emph{open quantum system} subjected to an external noise source. 

In the former approach (i), a quantum Markov chain can be described with a composition of \acp{tom} mapping a density operator $\rho$ to another density operator $\rho'$. \acp{tom} are matrices whose elements are \emph{completely positive maps} and whose column sums form a \emph{quantum operation} (for more details refer to \cref{sec:methods}). A special case of \acp{tom} are the \acp{tem} whose columns are discrete \acp{povm}. A quantum Markov chains can be, thereby, implemented by a sequence of quantum measurements~\cite{gudder2008}.

The second approach (ii) employs \acp{sse} to describe the physical quantum diffusion process. Given a system in the state $\rho(t)$, its stochastic evolution is determined by a \ac{sse} that takes the form $ \label{eq:sse} \dot{\rho}(t) = -i [H(t), \rho(t)],$ with $ \hbar = 1$, and where the Hamiltonian $ H(t) =  H_{s}(t) +H_{p}(t) $ consists of the sum of the Hamiltonian of the system $H_{s}(t)$ and the stochastic term $H_{p}(t)$ representing the stochastic dynamics to which the quantum system is subjected. Arbitrary sources of noise applied to optimally controlled quantum systems were very recently investigated with the \acp{sse} formalism by our group~\cite{muller2022information}.

\begin{figure*}[!ht]
\begin{subfigure}[b]{0.33\textwidth}
        \begin{center}
            \includegraphics[width=0.8\linewidth]{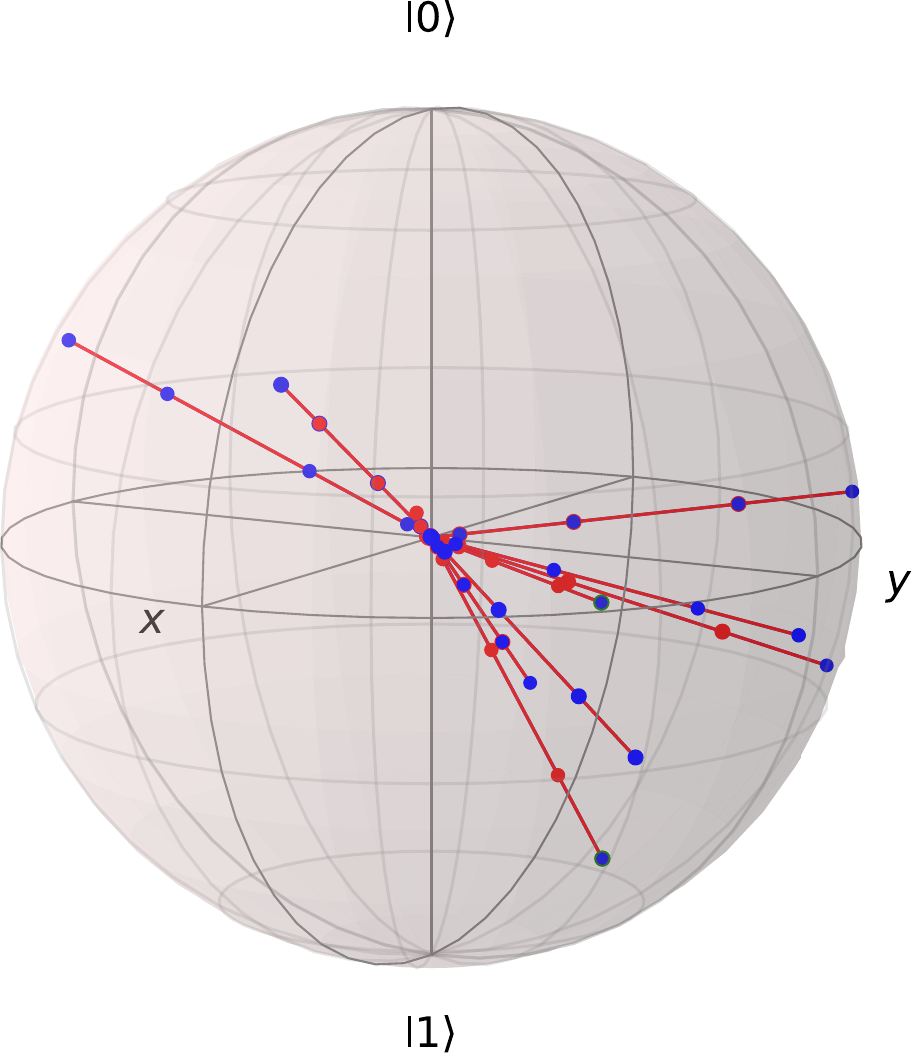}
            \caption{}
            \label{fig:fig3a}
        \end{center}
    \end{subfigure}%
    \begin{subfigure}[b]{0.33\textwidth}
        \begin{center}
            \includegraphics[width=0.8\linewidth]{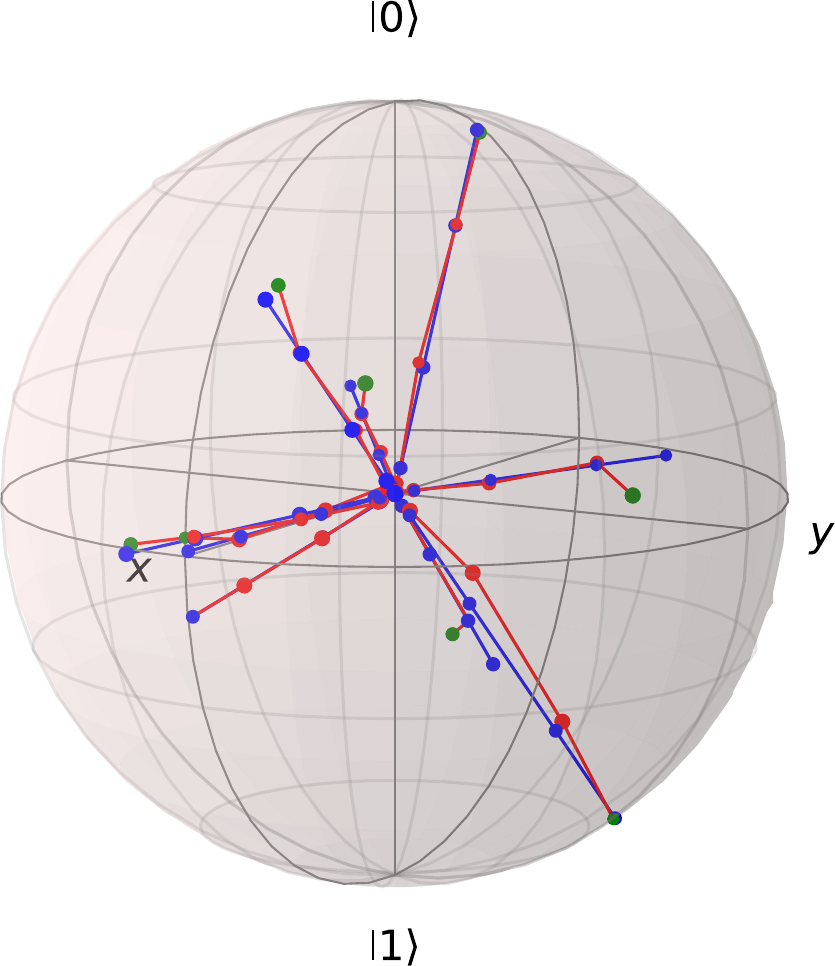}
            \caption{}
            \label{fig:fig3b}
        \end{center}
\end{subfigure}%
\begin{subfigure}[b]{0.33\textwidth}
    \includegraphics[width=\linewidth]{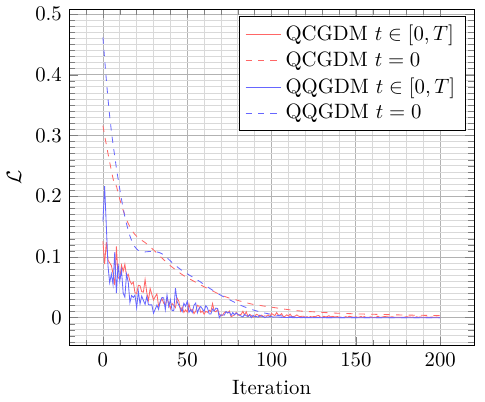}
            \caption{}
            \label{fig:fig3c}
\end{subfigure}%
\caption{Reconstruction of 10 random one-qubit pure states for trained (a) \acp{qcgdm} and (b) \acp{qqgdm}, and (c) the evolution of the quantum infidelity loss for both models for one of the 10 states. In (a) and (b), the blue segments represent the evolution of the forward dynamic implemented with a depolarizing quantum channel. In (a) the red segments are the evolutions of the backward process implemented with a neural network  and in (b) with a parameterized quantum circuit. The green points are the final reconstructed states. The solid lines in (c) report the evolution of the loss used during the training while the dashed lines are the same loss but calculated only on the final reconstructed states. The red lines refers to the \ac{qcgdm} and the blue lines to \ac{qqgdm}.}
\end{figure*}

As a practical implementation of the \ac{qcgdm} model, we decide to realize the forward dynamic as a discrete quantum Markovian chain composed by the iteration at each time step of a \emph{depolarizing} quantum channel $\rho_{t+1} = (1-p)\rho_t + p \frac{\mathbf{I}}{d}$ with $p \in [0, 1]$ and $t\in[0,T]$. In this way, the (quantum) information encoded in the initial quantum state $\rho_0$ is progressively degraded until we reach the maximally mixed state $\rho_T\equiv\frac{\mathbf{I}}{d}$ where $d$ is the dimension of the quantum system considered. Formally, the loss of information on the quantum state of a system can be quantified by the von Neumann entropy $S(\rho) = - \tr(\rho \log_2 \rho)$ that is zero for \emph{pure} states $\rho = \ketbra{\psi}{\psi}$, strictly positive for \emph{mixed} states $\rho = \sum_{i} p_{i} \ketbra{\psi_{i}}{\psi_{i}}$, with $p_{i} \geq 0$ and $\sum_{i} p_{i} = 1$, and it is maximal and equal to $S(\frac{I}{d}) = \log_2 d$ for the maximally mixed state~\cite{Nielsen2002}. Regarding the implementation of the backward, we use classical neural networks that predict $\hat{\rho}_t$ from the input at time $t+1$. A specific neural network is trained for each time step $t$ to maximize the fidelity between the $\rho_t$ obtained during the forward pass and the $\hat{\rho}_t$ obtained from the \ac{nn} when its input is $\rho_{t+1}$ in turn from the forward. In generation after the training, the \ac{nn} can be iteratively used to obtain $\hat{\rho}_0$ from the initial maximally mixed state $\frac{\mathbf{I}}{d}$ predicting, at each time step $t$, $\hat{\rho}_t$ from the previous prediction $\hat{\rho}_{t+1}$. In \cref{fig:fig3a} we show the simulation on a single qubit system of ten \acp{qcgdm} and $T=5$. The neural network is capable to reconstruct the information of the initial state $\rho_0$. The average reconstruction quantum fidelity on a sample of $100$ different random states is equal to $0.997 \pm 0.013$. In \cref{fig:fig3c} is visible, with the red curves, the evolution of the infidelity for batches of data during the training for one of the single states and also the infidelity loss between $\hat{\rho}_0$ and the $\rho_0$ in dashed (see \cref{sec:simulationQQ} for more details).

In our simulations, the forward process is classically simulated, therefore we obtain all the states $\rho_t$ directly from the simulator. If we want to use real \ac{nisq} devices, we need to perform quantum state tomography to obtain the $\rho_t$, and classically calculate the infidelity loss between the reconstructed state and the $\hat{\rho}_t$ predicted by the \ac{nn}. As a possible solution to avoid tomography, we could prepare the predicted $\hat{\rho}_t$ and use the \emph{swap test}, which is a common method to obtain the fidelity between quantum states~\cite{das2023quantum}, to compute the loss. In any case, both strategies are used only during the training of the model. After the training, the \ac{nn} is a generative model that classically simulate the generation of quantum states starting from the maximally mixed state, representing the noisy state.

The implementation of diffusion dynamics on quantum systems during the forward stage can allows the processing of the data information not only by classically simulated noise but also with quantum physical noise.
Here, as previously mentioned, let us remind that quantum noise is more general with respect to its classical counterpart. In particular, the noise distributions used in \acp{qcgdm} can be expressed (and more naturally arise by quantum dynamics) in more general and powerful forms respect to the typical Gaussian distributions that are commonly employed in classical \acp{dm}.
In this set up, at the end of the diffusion process, it is possible to obtain non-classical prior distributions related to entangled state that do not exist in the classical information scenario. In other terms there are probability density distributions that are purely quantum. This can be used to implement diffusion processes that are not possible to be implemented classically. At the end, during the denoising phase, classical \acp{nn} can be used in order to remove noise and thus finally generate new samples. Moreover, if the obtained prior distribution is not classical, it is possible to consider the adoption of the denoising \ac{nn} as a discriminator to identify probability distributions that are purely quantum. This could also be framed in a \emph{security} context. One can imagine a channel where the communication of data takes place with the application of a quantum diffusion process that maps to a purely quantum probability distribution. In that case, the receiver can restore and so obtain the initial information only with the training of a \ac{qnn} and thus only with a quantum device. This might be also exploited for quantum attacks/defence in cyber-security applications.

\subsection{\acf{qqgdm}} \label{sec:qqgdm}

In this last section we describe diffusion models within a fully quantum physical framework. Precisely, the training data, the diffusion process and the denoising process have all a quantum mechanical nature. This scenario can be obtained by exploiting the quantum tools described above, namely, quantum Markov chain or \ac{sse} for the forward diffusion phase, and a \ac{pqc} for the backward denoising phase. 

Accordingly, all the advantages described in \cref{sec:cqgdm,sec:qcgdm} hold. The adoption of a fully quantum pipeline for both the diffusion and denoising phases would allow the possibility to obtain purely quantum prior distributions that can be processed during the denoising phase with \acp{pqc} obtaining a generation process that is not feasible classically. As shown in \cref{fig:fig4}, the diffusion and denoising processes for \ac{qqgdm} are entirely located in the space of quantum distributions. This might lead to the speedup already described previously for \ac{cqgdm} and in addition to the possibility of exponentially reducing the computational resources for storing and processing of data information~\cite{yao2017}. Finally, it is also possible to access to complex quantum probability distributions that are impossible or much more difficult to treat classically.

Similarly to \ac{qcgdm} model, we propose a practical implementation of the forward quantum noise dynamic with a depolarizing channel to degrade the initial state $\rho_0$ to the maximally mixed state $\rho_T$.
The quantum backward process should be an equally noisy dynamic, in fact an unitary dynamics is not sufficient to reconstruct any initial state different from the maximally mixed one: $\frac{\mathbf{I}}{d}\mapsto U\frac{\mathbf{I}}{d}U^\dagger=\frac{\mathbf{I}}{d}$. For this reason, we propose an implementation of the backward via the interaction of a system with an external environment that we trace out at each time step. In detail, in our numerical simulations we consider a single-qubit system coupled via a parameterized circuit to another single-qubit that acts as the environment. During training, at each time step $t$, the system is initialized with $\rho_{t+1}$ and, after the environment tracing-out, the state is $\hat{\rho}_{t}$. In generation, it is possible to obtain $\hat{\rho}_0$, starting from the maximally mixed state $\frac{\mathbf{I}}{2}$, iteratively applying the denoising with the previous prediction as input. Similarly to the \ac{qcgdm}, the loss is based on the infidelity between the quantum states $\rho_t$ from the forward and $\hat{\rho}_t$ predicted during the backward (for more details see \cref{sec:simulationQQ}).
In our classical simulations we have access directly to the quantum states $\rho_t$ and $\hat{\rho}_t$ at each time step and, therefore, the loss can be straightforwardly computed. On real \ac{nisq} devices, the loss could be obtained with the help of swap test. Practically, the forward process could be implemented in two parallel copies of the system. The first one stops at time $t$ to prepare $\rho_t$ while the second one makes a further step to obtain $\rho_{t+1}$. The backward system can be appended to the second copy to prepare $\hat{\rho}_t$ and trained using the swap test between the two states.

In \cref{fig:fig3c} in blue is reported the evolution of the loss during the training of one \acp{qqgdm} and in \cref{fig:fig3b} the resulting simulations for ten different initial states. For each one, we train the model to reconstruct a random pure state with $T=5$. We can observe that the model is capable to learn the reverse dynamic from the maximally mixed state $\rho_T$ at the centre of the Bloch sphere to the state $\hat\rho_0$. Furthermore, we compute the average reconstruction quantum fidelity for $100$ random states obtaining $0.996 \pm 0.0086$.

\section{Conclusions}

The entanglement is a crucial quantum mechanical phenomenon occurring only in the quantum domain (not classical analogue) when two or more quantum systems interact. It is detected by measurement correlations between the quantum systems that cannot be described with classical physics. Accordingly, quantum systems are capable of representing distributions that are impossible to be produced efficiently with classical computers~\cite{bouland2019complexity,biamonte2017quantum}.
For this reason, a quantum diffusion process is capable to explore probability density functions that are not classically tractable. 

\begin{figure*}[!ht]
\begin{center}
\includegraphics[width=0.6\linewidth]{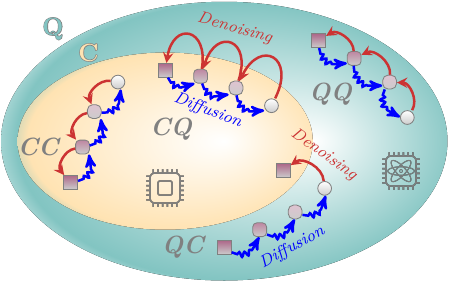}
\caption{Relationship between the space of the probability distributions that are tractable with classical computation ($\mathbf{C}$) and instead only with quantum computation ($\mathbf{Q}$). We show the trajectories arising from the mappings between probability distributions (colored and white shapes) during the diffusion (blue wavy arrows) and denoising (red arrows) processes for the four different combination: $\mathbf{CC}$, $\mathbf{CQ}$, $\mathbf{QC}$ and $\mathbf{CC}$ indicating whether the diffusion (first letter) and the denoising (second letter) are classical or quantum. The initial data distribution (squares) is progressively transformed during diffusion (changing color and shape) to an uninformative distribution represented by the white circles, and vice versa during denoising.
Completely classical models are limited to operate within the space of classically-tractable probability distributions, while completely quantum models can manipulate quantum-tractable probabilities. Models that have classical diffusion and quantum denoising are forced to work only with classical probabilities, but during the denoising phase they can exploit quantum properties within each step. Finally, models that have quantum diffusion and classical denoising can manipulate quantum probabilities during the forward, but in that case, it is not possible to train the classical backward to map those probability distributions.
}
\label{fig:fig4}
\end{center}
\end{figure*}
In \cref{fig:fig4} we highlight the relationship between the space of the probability distributions that are tractable with classic computers, which we denote with \emph{classical distributions} to be more concise, and the space of the probability distributions that are tractable with quantum devices, which we denote with \emph{quantum distributions} hereinafter. Moreover, we can observe several possible trajectories that map probability distributions to other probability distributions during the diffusion and denoising of the classical \ac{dm} and of the three proposed quantum approaches: \ac{cqgdm}, \ac{qcgdm} and \ac{qqgdm}. 

The classic \ac{dm} realizes maps from classical distributions to other classical distributions and the \ac{nn} that implements the denoising are trained to realize the inverse maps, i.e., to match the distributions crossed during the diffusion. 

In the \ac{cqgdm} approach, the diffusion process is implemented classically. Thus, all the probability distributions are necessarily classical. However, during the denoising process, the quantum dynamics is free to explore also the quantum probability space within each one of the steps hence exploiting potential (noise-assisted and/or quantum-enhanced) shortcuts. This may give advantages for the training of the denoising model. Moreover, when we evolve quantum systems within a \ac{qpu} it is possible to process and manipulate exponentially more information as compared to the classical case. 

When we consider the fully quantum framework \ac{qqgdm} we gain the advantage of exploring quantum distributions also during the diffusion phase. For this reason we could explore more complex noisy dynamics compared to the ones that can be simulated in classical computers. Moreover, the two processes can be experimentally implemented on real quantum processors. Furthermore, compared to the \ac{cqgdm} approach, and provided that the initial distribution of the dataset is quantum, it is possible to design a \ac{qqgdm} generative models that is capable of generating complex quantum data that are not analytically computable.

Besides, we would like point out that the \ac{qcgdm} approach can be challenging to be implemented. In detail, if the diffusion process leads to an entagled quantum distribution it is impossible, for the previously mentioned reasons, to efficiently train a classical \ac{nn} to perform the denoising. This context could be adopted as a proof of concept for the realization of a discriminator for the quantum distributions from the classical ones. In other words, if it is possible to train a model to perform the denoising, then the distribution is classical. 

After the second version of our work was published on arXiv~\cite{parigi2023quantumnoisedriven}, related works have appeared, showing indeed a remarkable research interest on
this topic. Compared to~\cite{parigi2023quantumnoisedriven}, the current version of our
manuscript contains two more examples for \ac{qcgdm} and \ac{qqgdm}.

More precisely, in Ref.~\cite{zhang2024generative} the authors propose an implementation of a quantum generative diffusion model where the forward phase is implemented with random unitaries iteratively applied at every time step $t$ and the backward with trainable \acp{pqc} with ancilla systems that are measured after each step. The main difference with respect to our approach is that in their case all the states $\rho_t$ and $\hat{\rho}_t$ during respectively the forward and backward phases are pure. Instead in our proposed method all the intermediate states, and in principle also the initial $\rho_0$ can be mixed states. 

Another research group in Ref.~\cite{cacioppo2023quantum} has implemented an hybrid model where the classic forward degrades images similarly to \acp{dm}, while the quantum backward reconstructs the images with \acp{pqc}. The main difference with our example implementation is that we use the \ac{pqc} as a real-valued function to predict the parameters of a classical Markov chain, while they use it as a quantum operator directly to reconstruct the state.

Finally, another recent work~\cite{chen2024quantum} has later proposed a fully quantum generative diffusion model where the forward phase iteratively degrades an initial quantum state $\rho_0$ using depolarizing channels to the maximally mixed state $\rho_T$ and the backward restores $\rho_0$ with a parameterized quantum operation. Our \ac{qqgdm} example approach is similar to their implementation, but with some differences. The first difference is that they use a cosine noise schedule in the forward, while we use a linear schedule. The second difference regards the denoising process where in order to reconstruct $\hat{\rho}_{t-1}$ from the state $\rho_{t}$ they also use a quantum circuit to perform the embedding of time $t$ on quantum state $\tau_t$ and then employ a \ac{pqc} that acts on the state of  total system $\tau_t \otimes \rho_t$. Next, they trace out the qubits of $\tau_t$ and obtain the state $\hat{\rho}_{t-1}$. Instead, in our implementation of the backward we use a single \ac{pqc} with parameters specific to each timestep $t$ without any embedding of time. Moreover, they also provide a second implementation that uses less resources (qubits) for the backward.

As a future outlook, we would like to realize the implementation of the \acp{qndgdm} either computationally via \ac{nisq} and/or physically by using quantum sensing technologies. In particular, regarding \acp{qcgdm} and \acp{qqgdm}, we propose to implement the diffusion process exploiting naturally noisy quantum dynamics in order to take advantage of the possible benefits of the quantum noise. Instead, regarding \acp{cqgdm} and \acp{qqgdm}, we propose to use quantum implemented \ac{qml} models, for instance \acp{qnn} and \acp{pqc}, to learn the denoising process. 

A possible future work direction could be to study the applicability of other kinds of loss functions. In fact, there are evidences that the adoption of \ac{kl} divergence loss in the context of quantum generative models leads to the formation of a new flavour of barren plateaus~\cite{rudolph2023trainability}. Related to this,
we plan to deepen the study of the possible noise-induced speedup of the diffusion dynamic and the trainability of the quantum \acp{qndgdm}. Moreover, other kinds of quantum channels could enhance the diffusion models by fully exploiting the quantum properties over all the diffusion and denoising dynamic. For instance, we can consider the adoption of coherent or non-unital noise such as the amplitude damping.

Finally, the design and realization of \acp{qndgdm}, with respect to classical \acp{dm}, could alleviate and reduce the computational resources (e.g. space of memory, time and energy) to successfully address ML applications such as generation of high-resolution images, the analysis and the prediction of rare events in time-series, and the learning of underlying patterns in experimental data coming also from very different fields as, among others, life and earth science, physics, quantum chemistry, medicine, material science, smart technology engineering, and finance.

\section{Methods}\label{sec:methods}
In this section we include some mathematical details on the classical and quantum tools for the diffusion and denoising processed discussed in the main text.

\subsection{Classical methods}\label{sec:classicalMethods}

Here we formalize the classical methods used in the standard generative diffusion models and for the relevant part of the proposed \ac{cqgdm} and \ac{qcgdm}. In particular, we consider classical Markov chains for a Gaussian pertubation and the \acp{nn} are used for the classical denoising.

The classical diffusion process~\cite{Sohl2015,Ho2020} starts from an initial data sample $\ve{x}_0$ drawn from an unknown generic distribution $p(\ve{x}_0)$. Gaussian noise is then iteratively injected for a number $T$ of time steps to degrade the data to $\ve{x}_T$ sampled from a prior Gaussian distribution $\mathcal{N}(0,\mathbf{I})$. In detail, the used Gaussian transition kernel is in the form:
\begin{equation}\label{eq:Gaussian_transition_kernel}
    p(\ve{x}_{t} | \ve{x}_{t-1}) = \mathcal{N}( \ve{x}_{t}; \sqrt{1-\beta_{t}} \ve{x}_{t-1}, \beta_{t}\mathbf{I}),
\end{equation}
where $\beta_{t} \in (0, 1)$ is an hyperparameter (fixed or scheduled over the time) for the model at the time step $t$ that describes the level of the injected noise, $\beta_{t}\mathbf{I}$ is the identity matrix, and $\ve{x}_{t}$ and $\ve{x}_{t-1}$ are the random variables at the time steps $t$ and $t-1$, respectively. In this way it is possible to calculate a tractable closed form for the trajectory:
\begin{equation}
    p(\ve{x}_{1:T}|\ve{x}_0)=\prod_{t=1}^T p(\ve{x}_{t} | \ve{x}_{t-1}).
\end{equation}
By obing so, for $T$ sufficiently high, $p(\ve{x}_{1:T}|\ve{x}_0)$ converges to an isotropic Gaussian $p(\ve{x}_T)\approx\mathcal{N}(0,\ve{I})$. Moreover, given an initial data $\ve{x}_0$ we can obtain a data sample $\ve{x}_{t}$ by sampling a Gaussian vector $\ve{\epsilon} \sim\mathcal{N}(0,\ve{I})$:
\begin{equation}\label{eq:data_samplig}
    \ve{x}_{t} = \sqrt{\bar{\alpha}_t}\ve{x}_{0} + \sqrt{1-\bar{\alpha}_t}\ve{\epsilon},
\end{equation}
where $\alpha_{t}:= 1-\beta_{t}$ and $\bar{\alpha}_t := \prod_{s=0}^{t}\alpha_{s}$.

The denoising phase starts from the Gaussian prior distribution and the transition kernel that is implemented is in the form:
\begin{equation}\label{eq:denoiseKernel}
    p_{\theta}(\ve{x}_{t-1} | \ve{x}_{t}) = \mathcal{N}( \ve{x}_{t-1}; \ve{\mu}_{\theta}(\ve{x}_{t}, t), \ve{\Sigma}_{\theta}(\ve{x}_{t}, t)),
\end{equation}
and the closed form for the trajectory is:
\begin{equation}
    p_\theta(\ve{x}_{0:T})=p(\ve{x}_T)\prod_{t=1}^T p_\theta(\ve{x}_{t-1} | \ve{x}_{t}).
\end{equation}
Usually, a \ac{nn}, specifically a U-Net architecture~\cite{Ronneberger2015}, is used to estimate the mean $\ve{\mu}_{\theta}(\ve{x}_{t}, t)$ and the covariance $\ve{\Sigma}_{\theta}(\ve{x}_{t}, t)$ in \cref{eq:denoiseKernel}. In principle, the approach to train the \ac{nn} would be to find the parameters $\theta$ such that $p_\theta(\ve{x}_0)$ would be maximized for each training sample $\ve{x}_0$. However $p_\theta(\ve{x}_0)$ is intractable because it is impossible to marginalize over all the possible trajectories. For this reason, the common approach is to minimize the \ac{kl} loss:
\begin{equation}\label{eq:kldiv}
\begin{split}
    \mathcal{L} & = \text{KL}(p(\ve{x}_{0:T})|| p_\theta(\ve{x}_{0:T}))\\
      & = - \mathbb{E}_p [\log p_\theta(\ve{x}_{0:T})] + \text{const} \\
      & = \mathbb{E} \Bigg[-\log p(\ve{x}_T) - \sum_{t\geq 1} \log \frac{p_{\theta}(\ve{x}_{t-1}|\ve{x}_{t})}{p(\ve{x}_{t} | \ve{x}_{t-1})}\Bigg]\\&\qquad\qquad\qquad\qquad\qquad\qquad\quad\quad\ \  + \text{const}\\
     & \geq \mathbb{E}[-\log p_{\theta}(\ve{x}_{0})] + \text{const}.
\end{split}    
\end{equation}

\subsection{Quantum methods}

Here we formalize the use of the quantum Markov chain introduced for the diffusion processes of \ac{qcgdm} and \ac{qqgdm} in \cref{sec:qcgdm,sec:qqgdm} and the \acp{qnn} used for the denoising of \ac{cqgdm} and \ac{qqgdm} in \cref{sec:cqgdm,sec:qqgdm}.

Formally, a quantum Markov chain can be described by two elements: i) a directed graph $G$ whose sites represent the possible state that the quantum system can occupy, ii) a \ac{tom} $\mathcal{E} = \mathcal{E}_{ij}$ whose elements are \emph{completely positive maps}~\cite{lindblad1975,Caruso2014} and whose column sums form a \emph{quantum operation}~\cite{gudder2008,Nielsen2002}. Formally, a positive a map is a linear transformation of one positive bounded operator into another. A completely positive map is a linear map $\phi: \mathcal{B(\mathcal{H})} \rightarrow \mathcal{B(\mathcal{H})}$, where $\mathcal{B(\mathcal{H})}$ is the set of bounded linear operators acting on the Hilbert space $\mathcal{H}$, such that the map $\phi \otimes I$ is positive on the space $\mathcal{B(\mathcal{H})} \otimes \mathcal{B(\mathcal{H'})}$ for any Hilbert space $\mathcal{H'}$.
A quantum operation is a completely positive map $\phi$ preserving the trace, i.e., $\tr(\rho) = \tr(\phi(\rho))$, with $\rho \in \mathcal{B(\mathcal{H})}$.
Physically, the elements $E_{ij}$ describe the passage operation of the quantum system from site $j$ to site $i$ in one time step. Given a density operator $\rho$, representing the state of system, the quantity $\mathcal{E}(\rho)$ is again a density operator. Moreover, if  $\mathcal{E}$ and $\mathcal{F}$ are two \acp{tom} with the same size and acting on the same Hilbert space, then the $\mathcal{EF}$ is again a \ac{tom} by matrix multiplication. Accordingly, the dynamics of the quantum system after a discrete number of time steps $n$ is described by the map $\mathcal{E}^{n} = \mathcal{E}\mathcal{E}^{n-1}$, with $n=2,3...$, and the initial state $\rho$ is transformed in the final state $\mathcal{E}^{n}(\rho)$.

Let us now introduce the concepts of \ac{qnn}~\cite{schuld2021machine} in the \ac{qml} framework and how they are trained.
Formally, a \ac{qnn} can be written as a product of layers of unitary operations:
\begin{equation}
    \hat{U}(\ve{\theta}) = \prod_{\ell=1}^{L} \hat{V}_\ell\hat{U}_\ell({\ve{\theta}_\ell}), 
\end{equation}
where $\hat{V}_\ell$ and $\hat{U}_\ell({\ve{\theta}_\ell})$ are fixed and parameterized unitary operations, respectively, for $\ell^{th}$ layer of \ac{qnn}. The output of the \ac{qnn} is:
\begin{equation}\label{eq:qnn}
    f(\ve{\theta})=\tr(\mathcal{M}\rho_\theta)
\end{equation}
where $\mathcal{M}$ is an Hermitian operator representing the physical observable, $\rho_\theta=\hat{U}(\ve{\theta}) \rho_0 \hat{U}^\dagger(\ve{\theta})$ and $\rho_0$ is the initial state, which is the input of the \ac{qnn}. The \ac{qnn} is optimized minimizing the difference between its output and the desired value. Generally, the latter is performed with the gradient descent method with the adoption of the parameters shift rule~\cite{Schuld2019evaluating}.

\subsection{Simulations}\label{sec:simulation}
Here we describe in detail both the model and its implementation regarding the simulation of the \ac{cqgdm}, \ac{qcgdm} and \ac{qqgdm} used to obtain the results of \cref{fig:fig2}, \cref{fig:fig3a} and \cref{fig:fig3b}, respectively. 

\subsubsection{\ac{cqgdm}}\label{sec:simulationCQ}
For the training of the \ac{cqgdm} model illustrated in \cref{sec:cqgdm}, we use a dataset composed of points $(x,y)$ distributed along a segment of the line $y=x$ in the interval $[-1,1]$. The model is trained on batches of $1\,000$ points uniformly random sampled for each step.

The diffusion process is implemented via a classical Markov chain composed of a sequence of Gaussian transition kernels as in \cref{eq:Gaussian_transition_kernel} in order to map the initial data distribution $p(\ve{x}_0)$ to an isotropic Gaussian $ p (\ve{x}_{T})$ with final time $T\equiv 40$. Furthermore, the data sampling at each time step $t$ is computed by using \cref{eq:data_samplig}.

The denoising process is realized via a \ac{pqc} and trained to estimate the mean $\ve{\mu}_{\theta}(\ve{x}_{t}, t)$ and the covariance $\ve{\Sigma}_{\theta}(\ve{x}_{t}, t)$ of the kernel of the classical denoising Markov chain in \cref{eq:denoiseKernel}. The model is built and simulated with the help of the \emph{Pennylane}~\cite{bergholm2022pennylane} and \emph{PyTorch}~\cite{paszke2019pytorch} libraries. More precisely, the \ac{pqc}, whose output is in the form of \cref{eq:qnn}, consists of a four qubits circuit divided in two concatenated parts called \emph{head} and \emph{tail}. The parameters of the head are shared among all the values of $t$, while the parameters of the tail are specific for each value of $t=0,\dots,39$. In particular, the head takes as input the values of the coordinates of a single point and encode them in the state of the first two qubits with an \emph{angle embedding}~\cite{schuld2021machine}, while the other two qubits are initialized to $\ket{0}$. After the embedding, the circuit is composed of $256$ layers of parametric rotations on the three axes for all the four qubits alternated by layers of entangling controlled not gates (in particular, we use \emph{StronglyEntanglingLayers} of \emph{Pennylane} with default arguments~\cite{schuld2020circuit}). At the end of the circuit, measurements are performed and the expectation values of the observable Pauli matrix $\sigma_z$ on all four qubits are computed. The tail is similarly composed, except that the first operation is the angle embedding of the four expectation values previously obtained from the head. In order to simplify the model, we assume that the denoising kernel is uncorrelated among the features and therefore, the covariance matrix is diagonal and only two values for the variance are necessary. This assumption, analogous to the one of the classical \ac{ddpm}, is justified by the fact that the denoising is implemented with a quantum-aided Markov chain and between each couple of time steps $t,t-1$ the data is classical and also the evolution of its distribution. Finally, the four expectation values measured from the tail are used for the predictions of the mean (the first two values) and variance (the second two values) of the classical kernel. In detail, we multiply the expectation values used for the mean by a factor $3$ in order to enlarge the possible range and the values for the variance are increased by $1$ to force positivity. The model is trained for $40\,000$ epochs on random batches of $1\,000$ points to minimize the \ac{kl} divergence loss of \cref{eq:kldiv} between the predicted and desired Gaussian distributions using \emph{Adam}~\cite{kingma2017adam} with learning rate $10^{-4}$. The evolution of such loss during the training is shown in \cref{fig:fig2b} averaged over every $1\,000$ iterations. The plots of \cref{fig:fig2a} are obtained, after the training of the model, using two different random batches of $1\,000$ points, one for the forward and another one for the backward. In detail, the top left plot in \cref{fig:fig2a} is the initial random batch of the data distributed with $p(\ve{x}_0)$ in the line segment. The batch is iteratively corrupted by the classical Markov chain with the kernel of \cref{eq:Gaussian_transition_kernel} and we report, on the top part of the figure, the evolution of the distribution at the intermediate time steps $t=8,16,24,32$ and at the final time $t=T\equiv40$. The kernel parameter in \cref{eq:Gaussian_transition_kernel} is taken equal to $\beta_t=0.3 \times \sigma(b_t)$ with $\sigma(x)=1/(1+e^{-x})$ the sigmoid function and $b_t$ equally spaced in $[-18, 10]$ for the 40 steps. The properties of the classical diffusion processes guarantee that for sufficiently large $T$ any initial distribution converges to the isotropic Gaussian distribution $p(\ve{x}_T)$. The bottom right plot is another random batch sampled from $\mathcal{N}(0,\ve{I})$ that is iteratively denoised with a parametrized classical Markov chain with the kernel of \cref{eq:denoiseKernel}. The kernel parameters are predicted by the previously mentioned \ac{pqc} and in the figure we report in the bottom part the evolution $p_\theta(\ve{x}_T),\dots,p_\theta(\ve{x}_{t:T}),\dots,p_\theta(\ve{x}_{0:T})$ of such process on the batch for the same values of $t$ of the forward.

\subsubsection{\ac{qcgdm}}\label{sec:simulationQC}
Here, we describe in details the implementation of the quantum forward and classical backward used for the single qubit \ac{qcgdm} model illustrated in \cref{sec:qcgdm}. The quantum forward dynamics is classically simulated in Pennylane with a circuit initialized with a pure quantum state that is iteratively degraded at each time step $t\in[1,T\equiv 5]$. The noise is introduced by a depolarizing channel applied $T$ times. Formally, the action of the channel on the quantum state is described by:
\begin{equation}
    \rho_{t} = (1-p_t)\rho_{t-1} + \frac{p_t}{3} (X\rho X+ Y\rho Y + Z\rho Z),
\end{equation}
where $p_t\in[0,1]$ represents the probability that the qubit depolarizes at time $t$. The values of $p_t$ are linearly scheduled in time with uniformly spaced values in $[1,T]$. All the values of the states $\rho_0,\dots,\rho_T$ are collected and used for the training of the backward phase.

The denoising process is implemented with neural networks that are trained to simulate the reverse noisy dynamic in order to obtain an approximation $\hat{\rho}_0$ of the initial pure state $\rho_0$ starting from the maximally mixed state $\frac{\mathbf{I}}{2}\equiv\rho_T$. The full process is implemented in PyTorch with $T$ different neural networks specialized to reconstruct the mixed state $\hat{\rho_t}$ from another mixed state at time $t+1$. Each one of them have a single hidden layer of $5$ neurons with ReLU activation function, $f(x)=max(x,0)$, for a total of 32 parameters, and it is trained to minimize the loss between the $\rho_t$ from the forward pass and the $\hat{\rho}_t$ predicted when the input is $\rho_{t+1}$. In detail, the loss $\mathcal{L}$ is computed and minimized for batches of $n\equiv 16$ reconstructed states $\hat{\rho}_t$ with random values of $t$. Formally, it is given by:
\begin{equation}\label{eq:loss}
    \mathcal{L} = \frac{1}{n}\sum_{t\sim[0,T]^n}\left(1-F(\hat{\rho}_t,\rho_t)\right),
\end{equation}
where $F(\rho_1, \rho_2) = \left( \tr \sqrt{\sqrt{\rho_1}\rho_2\sqrt{\rho_1}}\right)^2$ is the quantum fidelity between two states. The optimization is performed with Adam with learning rate $10^{-2}$. The loss function is classically calculated in our simulations. However, we can use the swap test in a possible implementation on \ac{nisq} devices.

As we formalized the problem, the inputs and outputs of the neural networks are mixed quantum states. A quantum state for one quit is represented by a $2\times 2$ positive semi-definite Hermitian matrix with trace one, for this reason to represent it are necessary only three real values. 
The three elements can be expressed as spherical coordinates with radius $r\in[0,1]$, polar angle $\theta\in[0,\pi]$ and azimuth $\phi\in[0,2\pi)$. Therefore, the input and output layers of the neural networks are both of dimension 3 and the final activation functions for the three neurons are respectively $\sigma(r)$, $\pi\sigma(\theta)$ and $2\pi\sigma(\phi)$ with $\sigma(\cdot)$ the sigmoid function.

\subsubsection{\ac{qqgdm}}\label{sec:simulationQQ}
In this section we give the details for the implementation of the \ac{qqgdm} in \cref{sec:qqgdm}. Here, the forward phase is implemented with the same depolarizing channel of \ac{qcgdm} explained in \cref{sec:simulationQC}, while the backward phase is realized with a quantum dynamic. The latter, is implemented with a non-unitary dynamic for each time step $t\in[T-1,0]$, with $T=5$. In particular, such dynamic is formalized in the context of open quantum systems with one qubit initialized with $\hat\rho_{t+1}$ (in prediction) that interacts with an environment that is traced out to prepare the output state $\hat{\rho}_t$ on the same system qubit. The environment relies on another qubit and the interaction with the system is realized with a trainable \ac{pqc} on both qubits. In detail, the used \acp{pqc} are implemented with the Pennylane StronglyEntanglingLayers with 5 layers and default other arguments for a total of 30 trainable parameters. The $t$-specific \acp{pqc} are trained analogously to the neural networks of \cref{sec:simulationQC}. In particular, using Adam with learning rate $10^{-2}$ to minimize the loss of \cref{eq:loss} calculated on batches composed by 16 pairs $(\hat{\rho}_{t},\rho_{t})$ with random values of $t$ where $\hat{\rho}_t$ is obtained when the system is initialized with $\rho_{t+1}$ from the forward.

\section*{Acknowledgements}
M.P. and S.M. acknowledge financial support from PNRR MUR project PE0000023-NQSTI. F.C. also acknowledges financial support by the European Commission’s Horizon Europe Framework Programme under the Research and Innovation Action GA n. 101070546–MUQUABIS, by the European Union’s Horizon 2020 research and innovation programme under FET-OPEN GA n. 828946–PATHOS, by the European Defence Agency under the project Q-LAMPS Contract No B PRJ-RT-989, and by the MUR Progetti di Ricerca di Rilevante Interesse Nazionale (PRIN) Bando 2022 - project n. 20227HSE83 – ThAI-MIA funded by the European Union-Next Generation EU.

\bibliographystyle{unsrt}
\bibliography{main}

\end{document}